\documentclass[%
 reprint,
 amsmath,amssymb,
 aps,
 prl,
floatfix,
]{revtex4-2}

\usepackage{graphicx}
\usepackage{epsfig} 
\usepackage{dcolumn}
\usepackage{bm}
\usepackage{epstopdf}
\usepackage{multirow}
\usepackage{amsmath}
\usepackage{booktabs}
\usepackage{color}
\usepackage{gensymb}
\usepackage{kantlipsum}  
\usepackage{hyperref}
\usepackage{ulem}
\usepackage{braket}
\usepackage{physics}
\usepackage[utf8]{inputenc}
\usepackage[T1]{fontenc}
\usepackage{mathptmx}

\begin{document}

\title{Ultracold Gas of Dipolar NaCs Ground State Molecules}

\preprint{APS/123-QED}

\author{Ian Stevenson}
\author{Aden Z.~Lam}
\author{Niccol\`{o} Bigagli}
\author{Claire Warner}
\author{Weijun Yuan}
\author{Siwei Zhang}
\author{Sebastian Will}\email{sebastian.will@columbia.edu}
\affiliation{%
  $^1$Department of Physics, Columbia University, New York, New York 10027, USA
}
\date{\today}

\begin{abstract}
We report on the creation of bosonic NaCs molecules in their absolute rovibrational ground state via stimulated Raman adiabatic passage.  We create ultracold gases with up to 22,000 dipolar NaCs molecules at a temperature of 300(50) nK and a peak density of $1.0(4) \times 10^{12}$ cm$^{-3}$. We demonstrate comprehensive quantum state control by preparing the molecules in a specific electronic, vibrational, rotational, and hyperfine state. Employing the tunability and strength of the permanent electric dipole moment of NaCs, we induce dipole moments of up to 2.6 D. Dipolar systems of NaCs molecules are uniquely suited to explore strongly interacting phases in dipolar quantum matter.
\end{abstract}

\maketitle

Ultracold dipolar molecules~\cite{carr2009cold, moses2017new} are a unique platform for the investigation of new areas in molecular and many-body quantum physics~\cite{lahaye2009physics,baranov2012condensed}. Studies with dipolar ground state molecules are enabling advances in quantum chemistry~\cite{krems2008cold,quemener2012ultracold, guo2018dipolar, hu2019direct} and have exciting prospects for quantum simulation~\cite{micheli2006toolbox,altman2021quantum}, quantum computing~\cite{demille2002quantum,andre2006coherent,sawant2020ultracold} and the investigation of novel quantum phases~\cite{buchler2007strongly,Capogrosso2010}. Their electric dipole moment gives rise to tunable long-range interactions that can be controlled via external electric \cite{de2011controlling, guo2018dipolar, bause2021collisions} or microwave~\cite{yan2013observation, will2016coherent, yan2020resonant} fields. Many applications of dipolar molecules will require quantum degenerate molecular gases. 

Molecular ensembles with the highest phase-space densities have been created via atom-by-atom assembly of ultracold molecules from ultracold gases of atoms~\cite{ni2008high,danzl2010ultracold}; first weakly bound Feshbach molecules~\cite{chin2010feshbach} are created and then transferred to their rovibrational ground state via stimulated Raman adiabatic passage (STIRAP)~\cite{vitanov2017stimulated}. Over the past years, this approach has been employed to create ground state molecules of KRb~\cite{ni2008high}, RbCs~\cite{takekoshi2014ultracold, molony2014creation}, NaK~\cite{park2015ultracold, seesselberg2018modeling, liu2019observation, voges2020ultracold}, NaRb~\cite{guo2016creation} and NaLi~\cite{rvachov2017long}. Recently, for the fermionic molecules KRb \cite{de2019degenerate} and NaK \cite{duda2021transition}, direct creation of degenerate molecular Fermi gases has been demonstrated. Also, collisional shielding \cite{valtolina2020dipolar, matsuda2020resonant,li2021tuning} and microwave shielding \cite{schindewolf2022evaporation} have been demonstrated to suppress two-body losses and enable evaporative cooling. For bosonic ground state molecules, collisional shielding, evaporative cooling, and the formation of a Bose-Einstein condensate are remaining outstanding goals.


\begin{figure} [t]
    \centering
    \includegraphics[width = 8.6 cm]{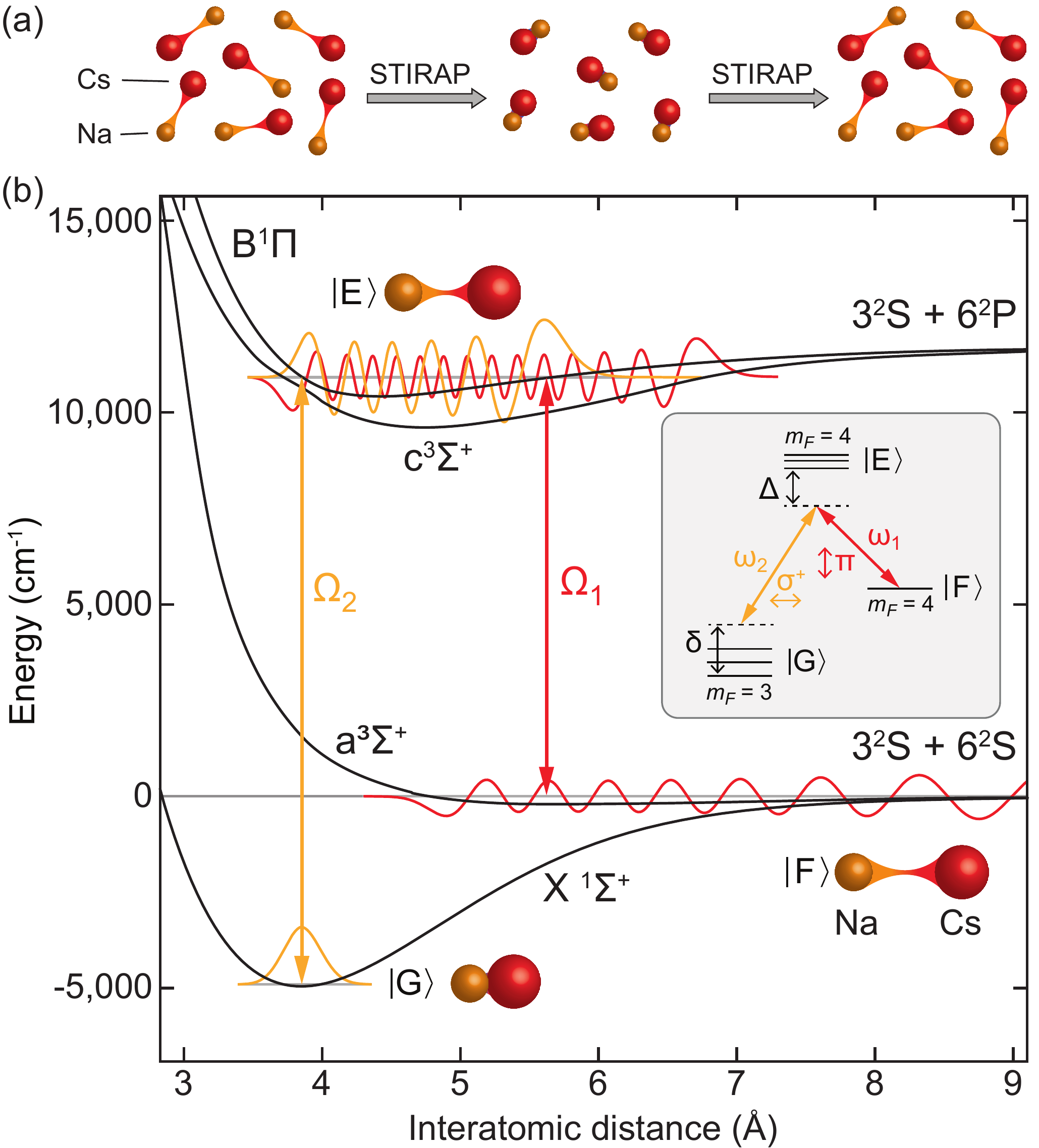}
    \caption{Two-photon pathway to the rovibrational ground state of NaCs. (a) An ultracold bulk gas of weakly bound NaCs Feshbach molecules is transferred to the ground state via STIRAP. Prior to detection, STIRAP is applied in reverse. (b) Molecular potentials of NaCs as relevant for transfer to the rovibrational ground state $X^1\Sigma^+$ $\ket{v=0, \ J=0, \ m_J=0}$. Here, $J$ denotes the total angular momentum and $m_J$ its projection on the quantization axis. $\Omega_1$ and $\Omega_2$ denote the Rabi frequencies of the Raman lasers. The inset shows the relevant sublevels, including the $m_F$ quantum numbers that are accessible with the given laser polarizations. $\Delta$ and $\delta$ denote the one- and two-photon detuning from the excited state and the ground state, respectively. Arrows indicate light polarizations relative to the quantization axis set by the vertical magnetic field.}
    \label{fig:1}
\end{figure}

In this Letter, we report on the production of ultracold gases of strongly dipolar NaCs molecules in their rovibrational ground state. Using STIRAP, we create up to 22,000 ground state molecules in a specific hyperfine state at a temperature of 300(50)~nK. In addition, we expose the molecular gases to electric fields of up to 2.1(2)~kV/cm and observe Stark shifts of the ground state that correspond to an induced dipole moment of up to 2.6~D, which is significantly larger than the highest induced dipole moment in previous work \cite{guo2018dipolar}. 

NaCs stands out due to its exceptionally large dipole moment of $d= 4.75(20)$ D~\cite{dagdigian1972molecular,aymar2005calculation} in the rovibrational ground state. As a result, the effective range of dipole-dipole interactions, $a_\mathrm{d} = md^2/(8 \pi \epsilon_0 \hbar^2)$, can reach tens of micrometers, an order of magnitude larger than for NaK and two orders of magnitude larger than for KRb ($m$ denotes the molecular mass). The large dipole moment also promises shielding techniques to be highly effective as they generally scale proportional to higher powers of $d$~\cite{julienne2011universal,martinez2017adimensional, lassabliere2018controlling}. For resonant collisional shielding in NaCs, a superb ratio of elastic to inelastic collisions of $10^6$ has been predicted \cite{martinez2017adimensional,li2021tuning}; for microwave shielding, the large dipole moment will facilitate the required strong microwave coupling between the ground and first excited rotational state \cite{karman2018microwave,anderegg2021observation, schindewolf2022evaporation}. In earlier work, ground state molecules of NaCs have been created via photoassociation in laser-cooled mixtures of Na and Cs~\cite{zabawa2011formation} and single NaCs ground state molecules have been created in optical tweezer traps~\cite{cairncross2021assembly}. Bulk systems of strongly dipolar NaCs molecules, which we create in this work, promise intriguing new avenues for quantum simulation and quantum information processing with dipolar molecules.

\begin{figure} [!t]
    \centering
    \includegraphics[width = 8.6 cm]{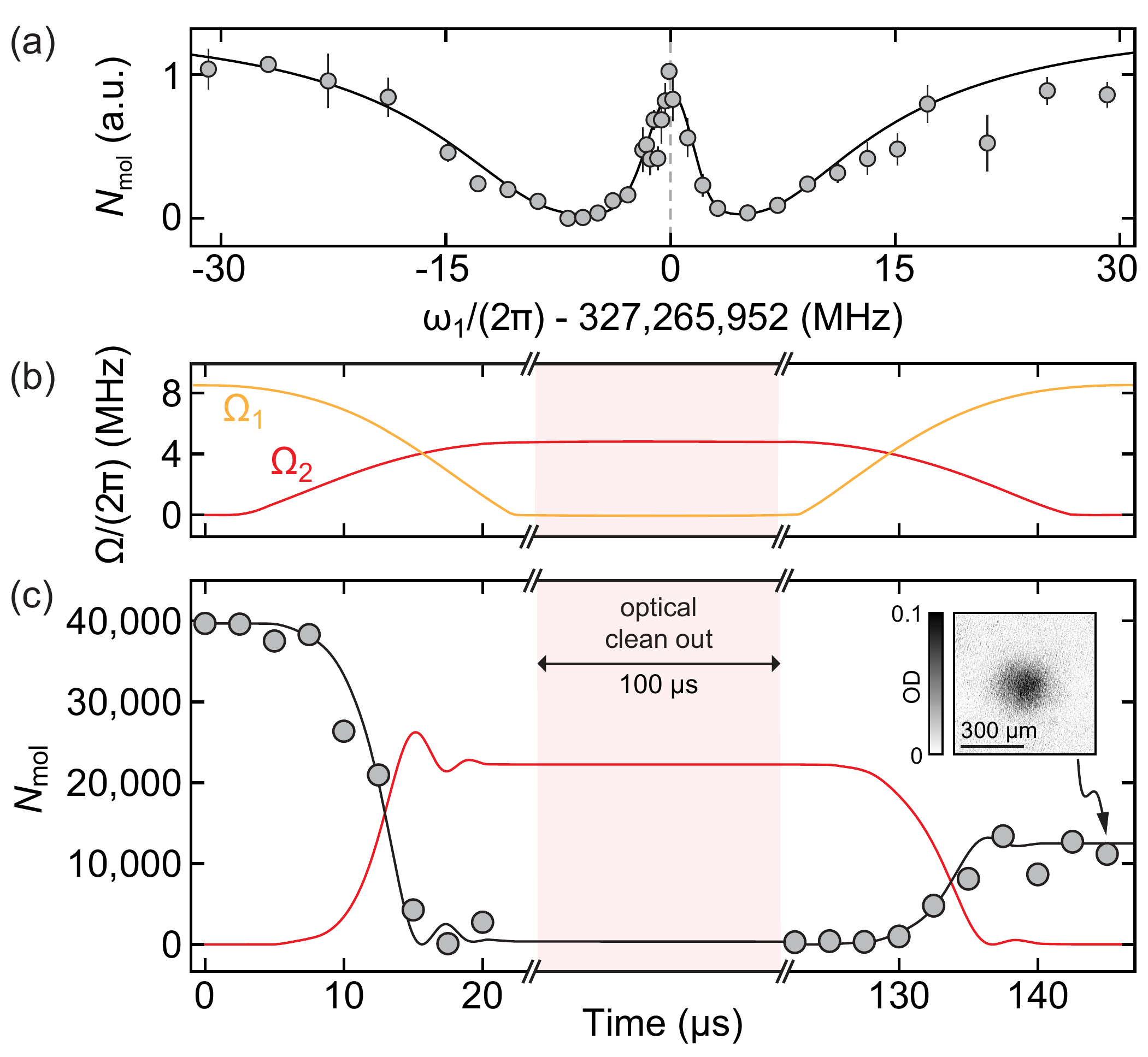}
    \caption{Creation of NaCs ground state molecules. (a) Coupling to the rovibrational ground state via $B^{1}\Pi \ket{v = 10}$ is observed in two-photon dark resonance spectroscopy. The solid line is a fit to a master equation model yielding Rabi frequencies $\Omega_1/2\pi = 0.6(1)$~MHz and   $\Omega_2/2\pi=8.8(5)$~MHz. (b) STIRAP transfer is realized via a ramp of laser intensities, giving rise to a dynamic change of the Rabi frequencies $\Omega_1$ (red) and $\Omega_2$ (orange) as shown. (c) Evolution of the number of Feshbach molecules (circles) during the STIRAP sequence. The black solid line shows a fit of the Feshbach molecule number to a master equation model that also yields the number of molecules in the ground state (red solid line). The inset shows a background-free absorption image of about $1.2\times10^4$ Feshbach molecules after a STIRAP round trip and 17 ms time of flight. For this data, $\Omega_1 / 2 \pi = 4.5$~MHz, $\Omega_2 / 2 \pi = 8.4$~MHz, $\Delta / 2 \pi = 90$~MHz, and $\delta / 2 \pi = 0$~MHz.}
    \label{fig:2}
\end{figure}

The experiment starts with an ultracold gas of $4 \times 10^4$ NaCs Fesh\-bach molecules. The molecules are associated from ultracold gases of Na and Cs via a magnetic field ramp across the Fesh\-bach resonance at 864.12(5)~G~\cite{warner2021overlapping,lam2022high}. Na and Cs are in their lowest hyperfine state $\ket{f, m_f} = \ket{1,1}$ and $\ket{3,3}$, respectively. The Fesh\-bach molecules are prepared at a magnetic field of 863.84(1)~G with a binding energy of about 500 kHz; their molecular state corresponds to $a^3\Sigma^+\ket{v=23}$ and is dominantly comprised of the spin triplet states $\ket{S=1, m_S=1 , m_{I_\mathrm{Na}} + m_{I_\mathrm{Cs}}=3}$ (about 70\%) and $\ket{S=1, m_S=0 , m_{I_\mathrm{Na}} + m_{I_\mathrm{Cs}}=4}$ (about 30\%). Here, $v$ denotes the vibrational state of the molecule, $S$ the quantum number of the electronic spin, and $m_S$ its projection on the quantization axis; $m_{I_\mathrm{Na}}$ ($m_{I_\mathrm{Cs}}$) is the projection of the Na (Cs) nuclear spin. The molecular gas is held in a crossed optical dipole trap operating at 1064 nm; the trap frequencies are $\omega=\{\omega_x, \omega_y, \omega_z\} = \{30(3),60(5),130(5)\}\,$Hz. For detection, the molecules are dissociated into free atoms via a reverse magnetic field ramp and imaged at low field using the procedure described in Ref.~\cite{lam2022high}.


We have developed a coherent two-photon scheme that transfers ultracold gases of NaCs Feshbach molecules to the rovibrational ground state via an electronically excited state that has mixed singlet and triplet spin character (see Fig.~\ref{fig:1}). For the laser that couples the Feshbach state $\ket{F}$ to the excited state $\ket{E}$ (laser 1), we use a titanium-sapphire laser that operates near 916 nm; its polarization is chosen vertical with respect to the quantization axis (defined by the vertical magnetic field). For the laser that couples the excited state $\ket{E}$ to the rovibrational ground state $\ket{G}$ (laser 2), we use a home-built extended cavity diode laser that operates near 632 nm; its polarization is chosen horizontal with respect to the quantization axis. In order to establish phase coherence between the lasers on the relevant time scale (less than 1 ms), we stabilize their absolute frequencies to the kilohertz-level by locking them to an optical cavity made from ultralow expansion glass via the Pound-Drever-Hall technique. The finesse of the cavity is about 22,000.  

As a bridge to the ground state, we use an electronically excited state in the $c^3\Sigma^+ \sim B^1\Pi$ complex as shown in Fig.~\ref{fig:1}(b). Due to strong spin-orbit coupling in NaCs \cite{korek2007theoretical,zaharova2009solution}, many vibrational states have strongly mixed singlet and triplet spin character. However, we have found that states of the $c^3\Sigma^+ \sim B^1\Pi$ complex have unconventionally broad linewidths, which is unsuitable for resonant or near-resonant STIRAP. For example, in recent work, Cairncross et al.~\cite{cairncross2021assembly} have demonstrated ground state transfer (without hyperfine selectivity) of an individually trapped NaCs molecule via far off-resonant Raman-Rabi transfer using the $c^3\Sigma^+\ket{v =26, J = 1}$ state. We found this state to have a linewidth of $2 \pi \times 51 (5)\,$MHz, which is significantly broader than realistically achievable Rabi frequencies. In this work, we have identified the  $B^{1}\Pi \ket{v = 10, \ J = 1, \ m_J = 1}$ as a suitable intermediate state. Additional spectroscopic details of the $c^3\Sigma^+ \sim B^1\Pi$ complex will be discussed in Ref.~\cite{warner2022excited}. 

We experimentally locate the ground state \cite{docenko2004spectroscopic} via dark resonance spectroscopy, as shown in Fig.~\ref{fig:2} (a). Dark resonance spectroscopy also allows us to measure the excited state linewidth $\Gamma$ and  Rabi couplings $\Omega_1$ and $\Omega_2$ by fitting the data with a three-state model~\cite{fleischhauer2005electromagnetically, bause2021efficient}. Combining the results of measurements at different laser powers, we obtain $\Gamma/2\pi=15(3)$~MHz for the $B^{1}\Pi \ket{v = 10, \ J = 1, \ m_J = 1}$ state. For the laser powers available in our setup, we can reach Rabi frequencies of up to $\Omega_1^\mathrm{max} / 2\pi \approx 12$~MHz and $\Omega_2^\mathrm{max} / 2\pi \approx 25$~MHz, which is on the order of the excited state linewidth. To realize STIRAP, we dynamically change the Rabi frequencies $\Omega_1$ and $\Omega_2$ [Fig.~\ref{fig:2} (b)], which transfers the molecules to the ground state and back, while the system remains in the dark state $\propto \Omega_2 |F\rangle + \Omega_1 |G\rangle$. Between forward and reverse STIRAP, a $100$ $\mu$s-pulse of resonant laser light removes remaining Na and Cs atoms and Fesh\-bach molecules. This increases the detected molecule number by a factor of three compared to a slower removal scheme based on a magnetic field gradient, which we employed in earlier work~\cite{lam2022high}.


\begin{figure} [b]
    \centering
    \includegraphics[width = 8.6 cm]{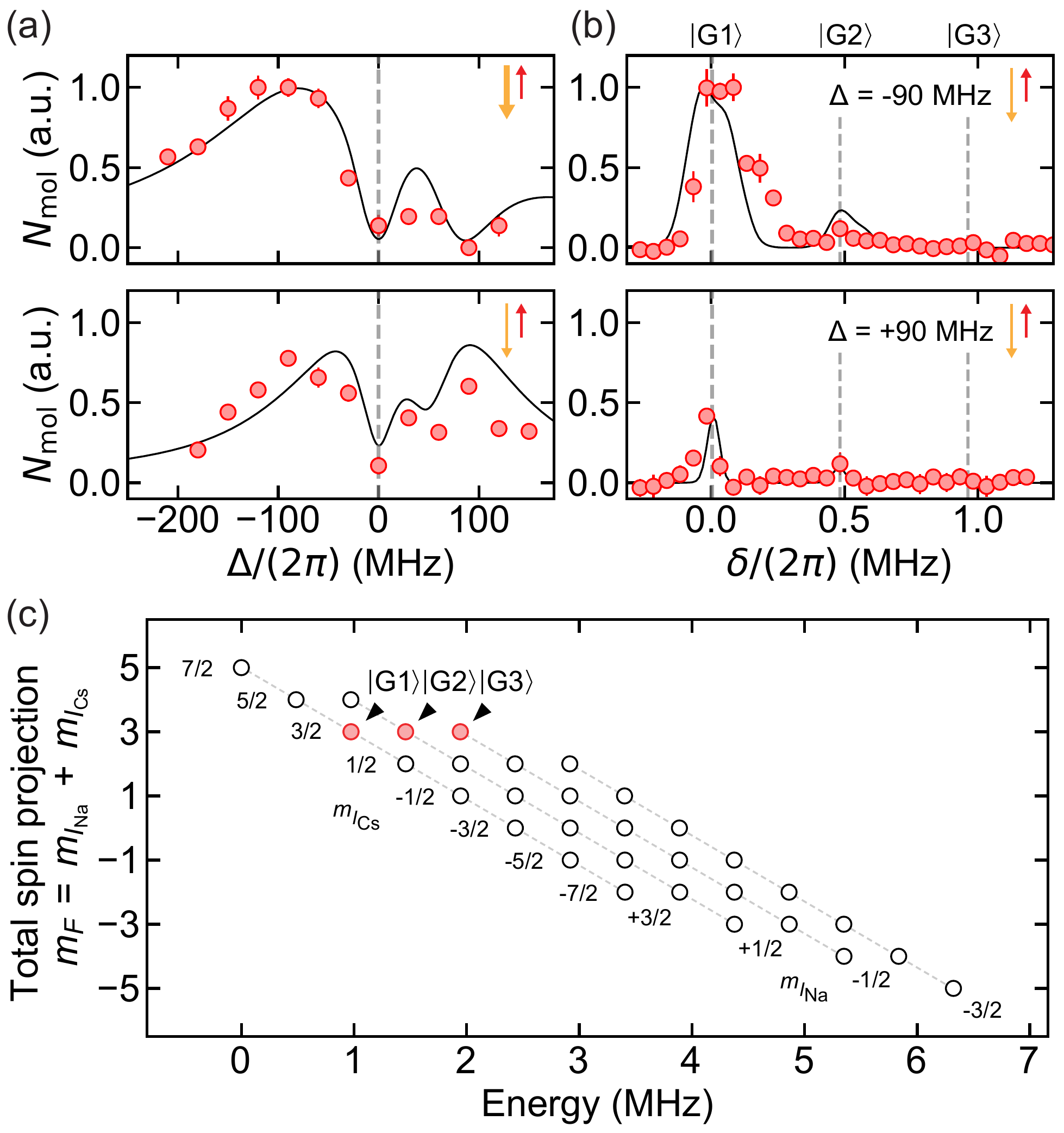}
    \caption{Controlling the hyperfine state of NaCs ground state molecules. (a) Dependence of transfer efficiency on the one-photon detuning $\Delta$, while staying on resonance with the ground state ($\delta=0$). Top (bottom) panel corresponds to $\Omega_1/2\pi = 4.5 $~MHz and $\Omega_2/2\pi = 8.4$~MHz ($\Omega_1/2\pi = 4.5$~MHz and $\Omega_2/2\pi = 4.5$~MHz). (b) Accessible hyperfine states for $\Delta/2\pi =-90$~MHz (top) and $\Delta/2\pi =+90$~MHz (bottom) with $\Omega_1 = \Omega_2 = 2\pi \times 4.5$~MHz. Solid lines in (a) and (b) are calculated from a master equation model to be reported in Ref.~\cite{warner2022excited}. (c) Energies of all $(2I_\mathrm{Na}+1)(2I_\mathrm{Cs}+1)=32$ hyperfine ground states at a magnetic field of 863.84~G. The hyperfine structure is in the Paschen-Back regime, dominated by the nuclear Zeeman effect, while scalar spin-spin interactions are weak ($c_4 = 3941.8$ Hz \cite{aldegunde2017hyperfine}). The initial Feshbach state has $m_F = 4$. Hyperfine states that can be addressed in our coupling scheme are shown as red dots.}
    \label{fig:3}
\end{figure}

Figure~\ref{fig:2} (c) shows transfer to the ground state by monitoring the Feshbach molecule population during the STIRAP sequence. We have found robust conditions for 20~$\mu$s long pulses and a one-photon detuning of $\Delta / 2 \pi = 90$~MHz. Starting with $4\times 10^4$ Feshbach molecules, we detect $1.2\times 10^4$ Feshbach molecules after the STIRAP round trip. This corresponds to a one-way transfer efficiency of 55(3)~\%. About $2.2\times 10^4$ ground-state molecules are created with a peak density of $1.0 (4) \times 10^{12}$~cm$^{-3}$.

We study the transfer efficiency as a function of one-photon detuning [see Fig.~\ref{fig:3} (a)]. We find that our system significantly deviates from the behavior of a canonical three-level STIRAP model, where most efficient transfer is expected on resonance ($\Delta = 0$) \cite{scully1999quantum}. Instead, our one-photon detuning spectra show nearly vanishing transfer efficiency on resonance for a pulse duration of 20~$\mu$s; in addition, they are highly asymmetric and tend to favor negative one-photon detuning. We attribute this behavior to unresolved hyperfine structure in the excited state; laser 1 with $\pi$-polarized light leaves $m_F = 4$ of the initial Feshbach state unchanged and can in principle couple to three hyperfine states fulfilling $m_F = 4$ [see inset of Fig.~\ref{fig:1} (b)]. Taking into account the multi-level structure, we have developed a numerical model \cite{warner2021overlapping}, which explains the behavior of our system in remarkable detail (solid lines in Fig.~\ref{fig:3}). The model suggests that the observed STIRAP efficiency is largely limited by the multi-level structure and not by laser noise. Remarkably, the model also predicts that efficient on-resonant STIRAP transfer becomes possible for very short pulses ($< 5$ $\mu$s), which we have confirmed experimentally. 


In Figure \ref{fig:3} (b) we demonstrate controlled transfer into a well-defined hyperfine level. To this end, we have measured distinct STIRAP resonances in the rovibrational ground state by varying the two-photon detuning $\delta$ for two different one-photon detunings $\Delta$. While laser 1 leaves $m_F = 4$ unchanged, laser 2 couples to the ground state via horizontal polarization, corresponding to $(\sigma^{+} + \sigma^{-})/\sqrt{2}$ light, which, in principle, gives access to $m_{F}=3$ and 5. However, experimentally we see dominant coupling into $|G1\rangle$ and weak coupling to $|G2\rangle$ with $m_{F}=3$; coupling to the absolute hyperfine ground state with $m_{F}=5$ is not observed and likely suppressed due to the absence of an $|m_{I_\mathrm{Cs}} = 7/2,  \ m_{I_\mathrm{Na}} = 3/2 \rangle$ admixture in $|E\rangle$. The observed splitting between $|G1\rangle$ and $|G2\rangle$ of 500(25) kHz is consistent with the prediction of the ground state hyperfine structure, shown in Fig.~\ref{fig:3} (c). 

\begin{figure} [t]
    \centering
    \includegraphics[width = 8.6 cm]{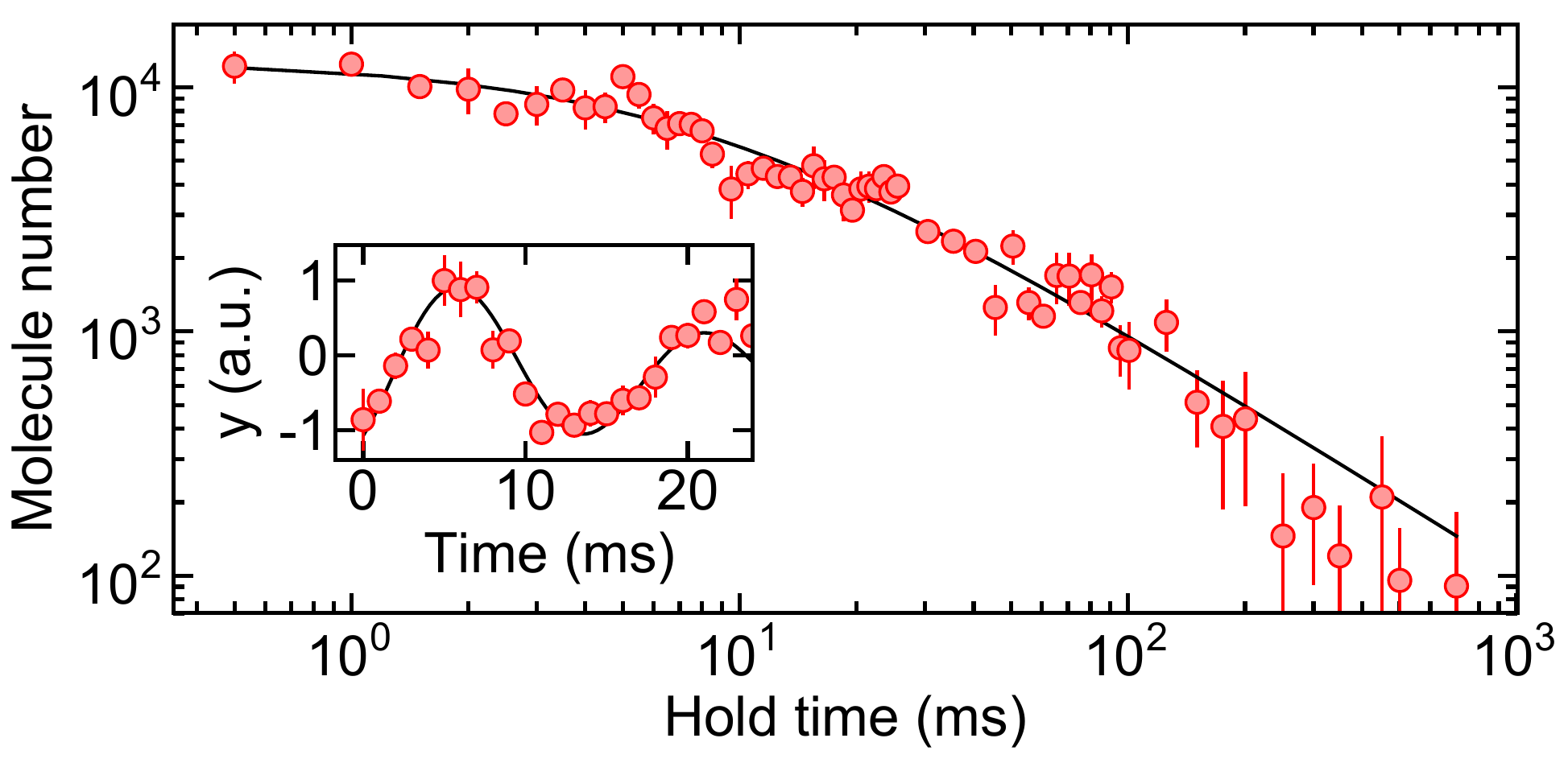}
    \caption{Lifetime of NaCs in the rovibrational ground state measured for an initial peak density of 5(3)$\times 10^{11}\,\mathrm{cm}^{-3}$ and temperature of 300(50)~nK. The solid line shows a fit to a two-body decay model. Data points correspond to the average of three experimental runs. The inset shows a trap frequency measurement of ground state molecules oscillating along the  $y$-axis. 
    From this we determine a ratio for the ac polarizabilities of ground state and Feshbach molecules of 0.79(2) in the optical dipole trap at 1064 nm.}
    \label{fig:4}
\end{figure}

\begin{figure} [t]
    \centering
    \includegraphics[width = 8.6 cm]{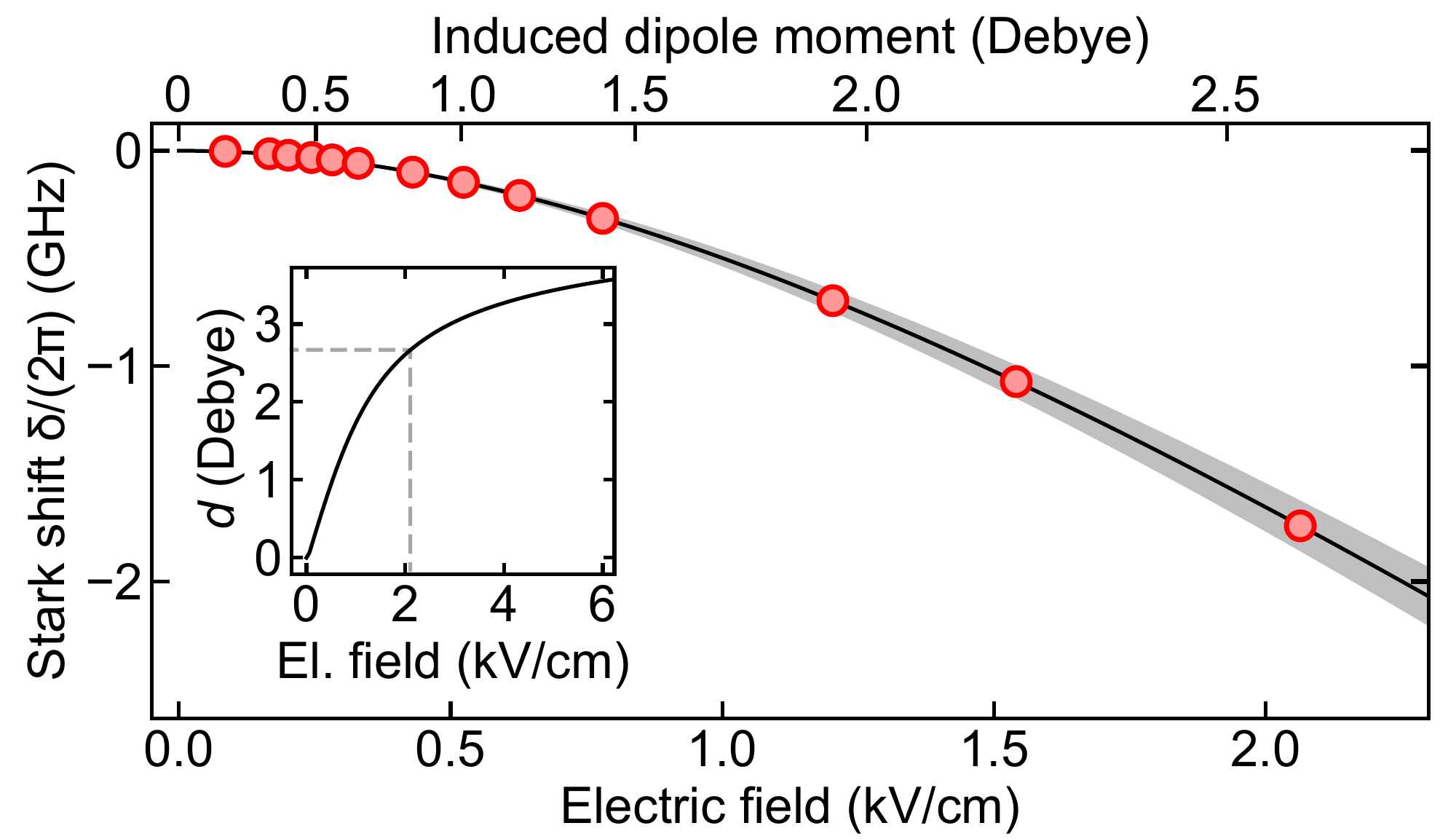}
    \caption{NaCs molecules in a homogeneous electric field. The Stark shift of the ground state is measured via dark resonance spectroscopy (red circles). Error bars are smaller than the point size. The solid line corresponds to the calculation of the expected Stark shift based on the known permanent dipole moment $d$ \cite{dagdigian1972molecular} and the rotational constant of the ground state. The gray shading reflects the uncertainty in the dipole moment. The Stark shift is used to calibrate the electric field strength. The inset shows the induced dipole moment $d$ versus dc electric field; the dashed rectangle indicates the dipole moments of up to 2.6 Debye reached in this work.}
    \label{fig:5}
\end{figure}


Equipped with the ability to create spin-polarized ultracold gases of NaCs ground state molecules, we study their collisional properties. Figure~\ref{fig:4} shows a lifetime measurement for an ensemble with a peak density of $n_0=5(3)\times 10^{11}$~cm$^{-3}$ and a temperature of $300(50)$~nK. The data closely follows a two-body loss model with a decay rate of $\Gamma_{2\mathrm{B}}^{\mathrm{exp}}=7(3)\times 10^{-10}$~cm$^3$~s$^{-1}$ and a characteristic decay time of $1/(\Gamma_\mathrm{2B} \bar{n}) = 10(1)\,$ms, where $\bar{n}$ is the average density \footnote{The fit function for two-body decay has the form $N(t) = N_0/(1 + \Gamma_\mathrm{2B} \bar{n} t)$, where $N_0$ is the initial molecule number. The average density $\bar{n}$ of a thermal gas in a harmonic trap is given by $\bar{n} = n_0/2^{3/2}$, where $n_0$ is the peak density.}. Our measurement agrees well with the predicted theoretical two-body collision rate, $\Gamma_{2\mathrm{B}}^{\mathrm{th}}=8\times 10^{-10} \mathrm{cm}^3~ \mathrm{s}^{-1}$~\cite{julienne2011universal}. This suggests that NaCs, similar to other bialkali species, exhibits unitary loss, although direct two-body loss is energetically forbidden~\cite{zuchowski2010reactions}. We will investigate the nature of the loss processes in NaCs in future work. This will be of critical interest given the plethora of open questions regarding the formation of collision complexes and loss processes in ultracold molecular gases~\cite{mayle2012statistical, mayle2013scattering}. While recent experiments with RbCs~\cite{gregory2020loss} and KRb~\cite{liu2020photo} have shown evidence for the formation of collision complexes that undergo loss after photoexcitation (e.g., via trap light), experiments with NaK~\cite{bause2021collisions} and NaRb~\cite{gersema2021probing} have not provided direct evidence for complex formation dynamics. NaCs will provide a valuable data point in this fast evolving field. 

Strong dipole-dipole interactions between NaCs molecules can be induced by aligning their electric dipole moments in the laboratory frame using a homogeneous electric field. In Figure~\ref{fig:5} we demonstrate the impact of a dc electric field on the ground state by monitoring its Stark shift that is directly related to the magnitude of the induced dipole moment. Homogeneous electric fields with up to 2.1 kV/cm are generated at the location of the molecules using indium tin oxide coated glass electrodes, mounted outside of the ultrahigh vacuum chamber. First, we confirm via one-photon spectroscopy that the Stark shift of the excited state is negligible even for the strongest fields. Then, we determine the Stark shift of the ground state via dark resonance spectroscopy. For the largest field, we observe a Stark shift of 1.8 GHz, which corresponds to an induced dipole moment of 2.6 Debye.  

The large dipole moment of NaCs promises to give access to strongly correlated phases in dipolar quantum systems, such as the formation of dipolar crystals in 2D bulk samples \cite{buchler2007strongly} or Mott insulators with fractional filling in molecular lattice gases \cite{Capogrosso2010}. In addition, it promises access to efficient mechanisms for the suppression of lossy two-body collisions. Recently, it has been shown that resonant collisional shielding~\cite{matsuda2020resonant,li2021tuning} and microwave shielding~\cite{anderegg2021observation,schindewolf2022evaporation} can enable effective evaporative cooling. For NaCs, resonant shielding is predicted to lead to an extremely favorable ratio of elastic to inelastic collisions of $10^6$~\cite{martinez2017adimensional,li2021tuning}. This requires accessing a F\"orster resonance in the $\ket{J, \ m_J} = \ket{1,\ 0}$ state, located at an electric field of 2.5 kV/cm~\cite{martinez2017adimensional}, which can be reached with minor technical upgrades. Also, microwave shielding is expected to be highly effective thanks to weak hyperfine interactions~\cite{karman2018microwave,lassabliere2018controlling}. This makes NaCs a promising candidate for evaporative cooling towards a Bose-Einstein condensate of dipolar ground state molecules, which is an important prerequisite for applications of dipolar molecules in quantum simulation \cite{baranov2012condensed}.


In conclusion, we have created an ultracold gas of 22,000 NaCs ground state molecules at 300(50) nK. Using STIRAP, we have prepared the molecules with full control over their electronic, vibrational, rotational, and hyperfine state. Demonstrating the tunability and strength of the electric dipole moment, we have exposed the molecular ensembles to electric fields that correspond to an induced dipole moment of up to 2.6 Debye. These conditions, combined with prospects for effective repulsive shielding protocols, make NaCs an exceptional candidate for the demonstration of evaporative cooling of bosonic ground state molecules, Bose-Einstein condensation, the exploration of new phases of matter, and applications in quantum simulation and quantum information. 

We are grateful to Eberhard Tiemann for providing a coupled-channel calculation for the Feshbach state and Kang-Kuen Ni and her group for fruitful discussions. We also thank Emily Bellingham for experimental assistance. This work was supported by an NSF CAREER Award (Award No.~1848466), an ONR DURIP Award (Award No.~N00014-21-1-2721) and a Lenfest Junior Faculty Development Grant from Columbia University. C.W.~acknowledges support from the Natural Sciences and Engineering Research Council of Canada (NSERC) and the Chien-Shiung Wu Family Foundation. W.Y.~acknowledges support from the Croucher Foundation. I.S.~was supported by the Ernest Kempton Adams Fund. S.W.~acknowledges additional support from the Alfred P. Sloan Foundation.

\bibliographystyle{apsrev4-1}

\end{document}